\documentclass[prl,superscriptaddress,showpacs, twocolumn,floatfix,amsmath,footinbib,amssymb]{revtex4-1}
\usepackage{amssymb}
\usepackage{amsmath}
\usepackage{epsfig}
\usepackage{t1enc}
\usepackage{soul}
\usepackage{color}
\newcommand{\beq}{\begin{equation}}
\newcommand{\eeq}{\end{equation}}
\newcommand{\bea}{\begin{eqnarray}}
\newcommand{\eea}{\end{eqnarray}}

\begin{document}

\title{Localized detection of quantum entanglement through the event horizon}
\author{Andrzej Dragan}
\affiliation{School of Mathematical Sciences, University of Nottingham, Nottingham NG7 2RD, United Kingdom}
\address{Institute of Theoretical Physics, University of Warsaw, Ho\.{z}a 69, 00-049 Warsaw, Poland}
\author{Jason Doukas}
\address{National Institute for Informatics, 2-1-2 Hitotsubashi, Chiyoda-ku, Tokyo 101-8430,  Japan}
\author{Eduardo Mart\'{i}n-Mart\'{i}nez}
\affiliation{Institute for Quantum Computing, University of Waterloo, 200 Univ. Avenue W, Waterloo, Ontario N2L 3G1, Canada}
\date{\today}
\begin{abstract}
We present a completely localized solution to the problem of entanglement degradation in non-inertial frames. A two mode squeezed state is considered from the viewpoint of two observers, Alice (inertial) and Rob (accelerated), and a model of localized projective detection is used to study the amount of entanglement that they are able to extract from the initial state. 
The Unruh vacuum noise plays only a minor role in the degradation process. The dominant source of degradation is a mode mismatch between the mode of the squeezed state Rob observes and the mode he is able to detect from his accelerated frame. Leakage of the initial mode through Rob's horizon places a limit on his ability to fully measure the state, leading to an inevitable degradation of entanglement that even in principle cannot be corrected by changing the hardware design of his detector.
\end{abstract}
\maketitle

\emph{--Introduction.\thinspace}
The semi-classical combination of gravity with quantum theory that lead to the apparent paradox of information loss in black holes \cite{Hawking1975}, naturally raises questions about the robustness of basic resources in quantum information theory in the presence of strong gravitational fields. Perhaps the most fundamental of these questions can be stated: ``How is the entanglement in the state of a quantum field affected by non-inertial motion of the observer?'' The pioneering works that attempted to answer this question involved non-localized states, usually plane waves \cite{Alsing2003, Fuentes2005} or Unruh modes \cite{Bruschi2010}  that spanned through the whole of spacetime. However, the observed entanglement degradation \cite{Bruschi2010},  was actually due to a particular parameterization of the initial state rather than the acceleration. The origin of this misinterpretation traces back to the inability to control the size of the global mode and the location of its observation, thus leaving the physical interpretation of the setting unclear. 

In this work we present  a completely localized solution to the question of entanglement degradation due to uniformly accelerated motion.  Our physical setup has a direct operational interpretation and the states and measurement devices used are localized both in space and time. Our approach leads to interesting new insights into the nature of the degradation process. For example we find, perhaps counter-intuitively, that for low accelerations the thermal Unruh noise is not the dominant cause of the degradation of entanglement. Rather, the entanglement loss is traced back to an inevitable mode mismatch that cannot be compensated for if the accelerated experimentalist has a device only sensitive to the modes defined in his co-moving reference frame. 

\emph{--Setup.\thinspace}
For simplicity we work with a massless 2D scalar field, which shares many of the same properties with light.
Two inertial observers Alice and Bob,  at rest with respect to each other, prepare an entangled two-mode squeezed vacuum state, $\hat{S}_{\text{AB}}|0\rangle_\text{M} = \exp\left[s\left(\hat{a}^\dagger\hat{b}^\dagger-\hat{a}\hat{b}\right)\right]|0\rangle_\text{M}$, where the annihilation  operators $\hat{a}$ and $\hat{b}$ are associated with two localized, orthogonal and spatially separated field modes $\phi_\text{A}(x,t)$ and $\phi_\text{B}(x,t)$ respectively. In order that these modes form well-defined annihilation operators we demand that the wavefunctions  be superpositions of positive Minkowski frequencies only.  Such states are commonly obtained via parametric down conversion in non-linear crystals \cite{PDC} and have been used, for example, in violations of CHSH inequality experiments \cite{Aspect}. 

In an inertial frame the entanglement in these states can be detected by projective measurements of quadratures carried out independently by Alice and Bob, resulting in an entanglement logarithmic negativity of $E_{\cal N} = 2s$. However, when one of the modes is measured by a relativistically accelerated observer, Rob,  the entanglement of the state changes due to a well-known relativistic transformation acting on the accelerated subsystem, effectively squeezing the the Minkowski vacuum state \cite{Unruh1976}. 

In order to study this non-inertial effect we will implicitly have in mind the model of a localized projective detector introduced in \cite{Dragan2012}. However, it will be sufficient in what follows to work at the abstract level of wavefunctions, where we will invariably use $\phi$ to denote the mode that the state is prepared in, and $\psi$ the mode that the detector is capable of measuring. In what follows we will use the Rindler coordinates $(c\tau,\xi)$ defined in  \cite{Dragan2012} to cover the part of Minkowski space accessible to Rob.

We suppose Alice (Rob) is in the possession of an inertial (accelerating) detector that couples to a mode of the field $\psi_{\text{A}}(x,t)$ ($\psi_{\text{B}}(\xi,\tau)$) with a corresponding annihilation operator $\hat{d}_\text{A}$ ($\hat{d}_\text{B}$), again demanding that the mode associated with this detector be a localized superposition of positive frequency Minkowski (Rindler) plane waves.  
%
We would like to focus on the effect accelerated motion has on Rob's ability to measure the part of the initial state accessible to Bob. 
The effective state accessible to Alice and Rob will be the mixed Gaussian state formed by tracing out all the modes orthogonal to $\psi_{\text{A}}$ and $\psi_{\text{B}}$. Any Gaussian state is fully characterized by its covariance matrix $\sigma$. To determine it in practice, an ensemble of states must be prepared and a series of projective measurements of quadrature operators must be carried out on consecutive copies. We assume this is done at $t=0$ when Rob's velocity is zero and doppler-shift effects \cite{Downes2012} do not arise.  It is enough to measure two orthogonal quadratures of each mode: $\hat{x}_k = \frac{1}{\sqrt{2}}\left(\hat{d}_k + \hat{d}_k^\dagger\right)$ and $\hat{p}_k = \frac{1}{\sqrt{2}i}\left(\hat{d}_k - \hat{d}_k^\dagger\right)$, where $k\in\{\text{A},\text{B}\}$ and construct the correlations between the measurement outcomes: $\sigma_{ij} = \langle \hat{X}_i \hat{X}_j+ \hat{X}_j \hat{X}_i\rangle$, where $\boldsymbol{\hat{X}} = (\hat{x}_\text{A}, \hat{p}_\text{A}, \hat{x}_\text{B}, \hat{p}_\text{B})$. 

In order to calculate the covariance matrix, we need to evaluate the operators $\hat{S}^{\dagger}_{\text{AB}}\hat{d}_k\hat{S}_{\text{AB}}$. This can be done by trivially rewriting $\hat{d}_\text{A} = (\psi_\text{A},\phi_\text{A})\hat{a} + \hat{d}_\text{A} - (\psi_\text{A},\phi_\text{A})\hat{a}$, where $(\cdot,\cdot)$ is the Klein-Gordon inner product, and noting that the part,  $\hat{d}_\text{A} - (\psi_\text{A},\phi_\text{A})\hat{a}$, does not contain $\hat{a}$ operators. A similar decomposition can be introduced for the  $\hat{d}_\text{B}$ operator (taking out $\hat{b}$ and $\hat{b}^{\dagger}$) leading to the following commutation relations:
\begin{eqnarray}
\hat{S}^\dagger_{\text{AB}}\hat{d}_\text{A}\hat{S}_{\text{AB}} &=& \alpha(\cosh s\, \hat{a} - \sinh s\, \hat{b}^\dagger) +\hat{d}_\text{A}-\alpha\hat{a},
\\ 
\hat{S}^\dagger_{\text{AB}}\hat{d}_\text{B}\hat{S}_{\text{AB}} &=& \beta(\cosh s\, \hat{b} - \sinh s\, \hat{a}^\dagger)+\hat{d}_\text{B}-\beta\hat{b}-\beta'\hat{b}^\dagger\nonumber\\
&~&+\beta'(\cosh s\, \hat{b}^\dagger - \sinh s\, \hat{a}),
\end{eqnarray}
where $\alpha\equiv(\psi_\text{A},\phi_\text{A})$, $\beta\equiv(\psi_\text{B},\phi_\text{B})$, and $\beta'\equiv(\psi_\text{B},\phi^\star_\text{B})$. With these relations and the fact that the operators $\hat{a}$, $\hat{b}$ and $\hat{d}_{\text{A}}$ annihilate the vacuum $|0\rangle_\text{M}$ we obtain the covariance matrix of the state $\hat{S}_{\text{AB}}|0\rangle_\text{M}$:
\begin{widetext}
\begin{eqnarray}
\label{covariance}
\sigma &=&
\openone+2\langle\hat{n}\rangle_U
\begin{pmatrix}
0&0&0&0\\
0&0&0&0\\
0&0&1&0 \\
0&0&0&1
\end{pmatrix}+2\sinh^2 s
\begin{pmatrix}
|\alpha|^2&0&0&0\\
0&|\alpha|^2&0&0\\
0&0&|\beta+\beta'^\star|^2&2\,\text{Im}(\beta\beta') \\
0&0&2\,\text{Im}(\beta\beta')&|\beta-\beta'^\star|^2
\end{pmatrix}\nonumber \\ \nonumber \\ 
& &+ 
\sinh 2s\begin{pmatrix}
0&0&-\text{Re}[\alpha(\beta+\beta'^\star)]&-\text{Im}[\alpha(\beta-\beta'^\star)]\\
0&0&-\text{Im}[\alpha(\beta+\beta'^\star)]&\text{Re}[\alpha(\beta-\beta'^\star)]\\
-\text{Re}[\alpha(\beta+\beta'^\star)]&-\text{Im}[\alpha(\beta+\beta'^\star)]&0&0 \\
-\text{Im}[\alpha(\beta-\beta'^\star)]&\text{Re}[\alpha(\beta-\beta'^\star)]&0&0
\end{pmatrix},
\end{eqnarray}
\end{widetext}
where
\begin{equation}
\label{noise}
\langle\hat{n}\rangle_U =  \sum_{k} \frac{|(\psi_\text{B},w_{Ik})|^2}{e^{\frac{2\pi |k| c^2}{a}}-1}; \quad w_{Ik}=\frac{1}{\sqrt{4\pi |k| c}}e^{i(k\xi-|k|c\tau)},
\end{equation}
is the average number of Unruh particles seen by an accelerated detector in the vacuum \cite{Dragan2012}. The appearance of thermal noise from the vacuum is a well known consequence of accelerated motion in the vacuum \cite{Unruh1976} and we shall strictly refer to it as {\it Unruh noise}. 

When the squeezing parameter $s\rightarrow0$, the state becomes the usual Minkowski vacuum $|0\rangle_\text{M}$ and, unlike Unruh-DeWitt detectors \cite{Reznik2005}, we find that no correlations are present in the covariance matrix \eqref{covariance}.  This situation should also be distinguished from when both detectors are counter-accelerating with equal magnitude, in that case correlations do exist \cite{Dragan2012}. 

\emph{--Entanglement degradation.\thinspace}
In order to quantify the non-local correlations measured by Alice and Rob we use the logarithmic negativity, which constitutes an upper bound to the distillable entanglement. It can be completely calculated from the elements of the covariance matrix \eqref{covariance} \cite{Adesso2004}:
\begin{eqnarray}
\label{negativity}
E_{\cal N} ={\rm{Max}}\left[ 0,-\log\sqrt{\frac{\Delta-\sqrt{\Delta^2-4\det\sigma}}{2}}\right],
\end{eqnarray}
where $\Delta = \sigma_{11}\sigma_{22}-\sigma_{12}^2 + \sigma_{33}\sigma_{44}-\sigma_{34}^2-2\sigma_{13}\sigma_{24}+2\sigma_{14}\sigma_{23}$. 

When Rob's proper acceleration vanishes and the modes $\phi_k$ and $\psi_k$ are chosen to match perfectly, $\alpha=\beta=1$, $\langle\hat{n}\rangle_U=\beta'=0$ then $E_{\cal N}=2s$. For non-zero proper accelerations the expressions \eqref{covariance} and \eqref{noise} can be substituted into \eqref{negativity} and studied numerically. In order to perform this calculation, specific forms of $\phi_\text{A}(x,t)$, $\phi_\text{B}(x,t)$, $\psi_\text{A}(x,t)$ and $\psi_\text{B}(\xi,\tau)$ need to be chosen. Since we consider measurements carried out at $t=\tau=0$, it is sufficient to specify the modes by their wavefunctions and their first derivatives at this time only. Since Alice is inertial we assume the idealized scenario when her detector couples perfectly to her mode of the entangled state i.e., $\alpha=1$. Any degradation of entanglement will therefore be a consequence of the specification of Rob's detector. 

We model the functional shape of Bob's initial state mode with the Gaussian:
\begin{eqnarray}
\label{modeAfunction}
\phi_\text{B}(x,0) &=& \frac{1}{\sqrt{N\sqrt{2\pi}}}\exp\left[-\frac{x^2}{L^2} + i \frac{N}{L}x\right],\nonumber \\ \nonumber \\
\partial_t{\phi}_\text{B}(x,0) &=& -i\frac{Nc}{L}{\phi}_\text{B}(x,0),
\end{eqnarray}
but modify it with a low frequency filter, eliminating all frequencies characterized by $k<\tfrac{1}{3L}$. This removes an unphysical divergence in the spectrum at zero wave number but does not appreciably change the shape or localization of the mode near $t=0$. Here, $N$ is the characteristic frequency about which the mode is centered. For convenience we take it as a large natural number ($>3$) which ensures the component of negative Minkowski frequency plane waves present in $\psi_\text{B}$ is negligible. $L$ is the spacial width of the localized mode and in combination with Rob's acceleration the dimensionless quantity, $\tfrac{a L}{c^2}$, will set the scale at which entanglement degradation effects become important. 

For each acceleration Rob is given an identical initial state. However, his position, $x(0)=\frac{c^2}{a}$, is acceleration dependent, and therefore the initial mode must be translated according to $\phi_\text{B}(x,0)\rightarrow \phi_\text{B}(x-\tfrac{1}{a},0)$. It is important to realize that such repositioning does not change the initial state, it is merely a computational convenience allowing us to describe each acceleration using a single Rindler coordinate chart. Alternatively, the initial state could be kept fixed and the origin of the Rindler coordinate chart could be adjusted such that Rob passes through the centre of the mode for each acceleration.
 
Having chosen a localized initial state, there remains the question of deciding which mode Rob's detector should couple to. In practice, the response of Rob's detector would be dependent on its physical design and it may even depend on his acceleration. As mentioned, inertial observers like Alice can always make the idealized detector assumption effectively setting $\alpha=1$, but in Rob's case such a setting is immediately ruled out by the existence of negative Rindler frequencies in $\phi_\text{B}$.  Thus some alternative choice must be made.

Consider constructing Rob's detector's mode in the Rindler frame in analogy to the construction of Bob's mode (\ref{modeAfunction}) in the inertial frame. 
Apart from the spacial translation, the mode $\psi_B$ is obtained from the mode of a resting detector by replacing the spacial coordinate $x$ with the conformal Rindler coordinate $\xi$ and replacing $L$ with the appropriate length in the conformal coordinates, $\widetilde{L}=\frac{2c^2}{a}\text{asinh}\left(\frac{aL}{2c^2}\right)$. As a result,
\begin{eqnarray}
\label{modeBfunction}
{\psi}_\text{B}(\xi,0) &=& \frac{1}{\sqrt{N\sqrt{2\pi}}}\exp\left[-\frac{\xi^2}{\widetilde{L}^2} + i \frac{N}{\widetilde{L}}\xi\right],\nonumber \\ \nonumber \\
\partial_\tau{\psi}_\text{B}(\xi,0) &=& -i\frac{Nc}{\widetilde{L}}{\psi}_\text{B}(\xi,0).
\end{eqnarray}
Again we assume a high pass filter and take $N$ large, producing an annihilation mode with the correct properties in the Rindler frame. This construction can be physically motivated by appealing to the transformation that undergoes an ideal cavity mode when the cavity walls are accelerated \cite{Dragan2012}.  

\begin{figure}[h]
\begin{center}
\includegraphics[width=\columnwidth]{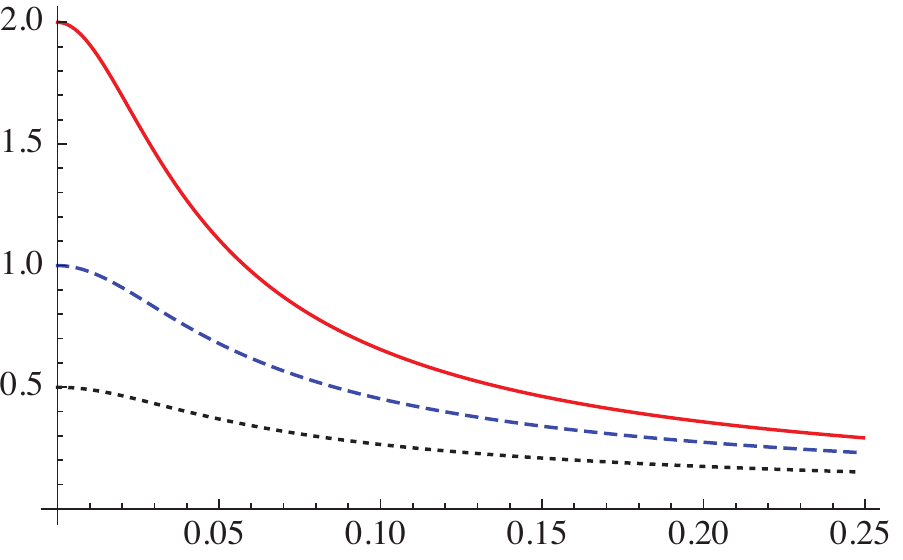}
\end{center}
\caption{\label{entanglementSs}Logarithmic negativity $E_{\cal N}$ as a function of the dimensionless parameter $\frac{aL}{c^2}$ for $N=100$ and the squeezing parameter $s=1$ (solid line), $s=0.5$ (dashed line) and $s=0.25$ (dotted line).  }
\end{figure}

To ensure that the acceleration is approximately uniform over the length of the detector, we also assume that $\frac{a \widetilde{L}}{c^2}\ll1$, or equivalently that $\widetilde{L}\approx L$.  In this limit, using \eqref{modeAfunction} and \eqref{modeBfunction} we obtain analytic estimates for the scalar products appearing in the covariance matrix:
\begin{equation}
|\beta|\approx \left(1+\left(\frac{NaL}{4c^2}\right)^2\right)^{-1/4}, 
\end{equation}
and $\beta' \approx0$.  Since $|\beta|<1$ for non-zero accelerations, a component of the degradation of entanglement will come from a mode mismatch between Rob's detector and the mode $\phi_{\text{B}}$ that he observes.  Indeed, in the large $N$ limit $\beta\rightarrow0$ and this mismatch leads to a complete degradation of the entanglement even for small accelerations. One may wonder how this source of disentanglement compares with that coming from the ambient Unruh noise (\ref{noise}). The expected number of Unruh particles per Rindler frequency satisfies a Bose-Einstein distribution at the Unruh temperature, 
\begin{equation}\label{eqn:VacParticleExpValue}
\langle \hat{n}_{k}\rangle=\frac{1}{e^{2 \pi |k| c^2/a}-1}.
\end{equation}  
These particles are mostly populated in frequencies below a critical value $k_c=\frac{a}{2 \pi c^2}$.  However, the assumed low frequency filter in the spectrum of (\ref{modeBfunction}) and the assumption of a small acceleration spread 
 over the detector together imply that the frequencies Rob's detector is sensitive to are greater than the critical value, $k>k_c$. Therefore virtually all the thermal particles are undetectable, $\langle n_U\rangle \sim0$. This is confirmed numerically.  Thus, the mode mismatch is truly the dominant source of entanglement degradation in these localized models. In Fig.~\ref{entanglementSs} we plot the numerically calculated entanglement as a function of the dimensionless parameter $\frac{aL}{c^2}$ for several values of initial squeezing. For all values of the initial squeezing, the entanglement approaches zero as $\frac{aL}{c^2}$ is increased.

This simple example illustrates the key features of the degradation of entanglement that are present generally for other detector mode shapes. Yet one may be left wondering if the degradation effects are truly fundamental to the acceleration or are merely a consequence of choosing a bad detector. In particular, could Rob completely eliminate all of the mode mismatch by cleverly redefining the mode that his detector responds to at each acceleration? 
In what follows we will consider an alternative definition of the detector mode, defining it to be the best detector at each acceleration.  

\begin{figure}[h]
\begin{center}
\includegraphics[scale=0.9]{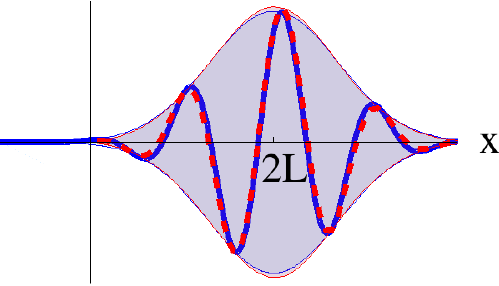}\includegraphics[scale=0.9]{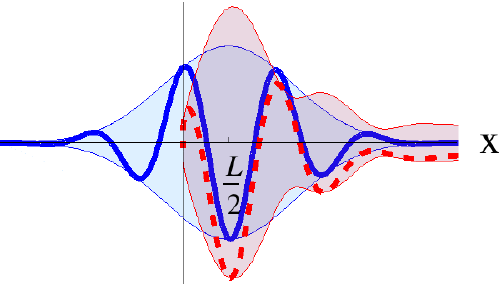}
\end{center}
\caption{\label{fig:optmodepic} Visualization of the optimized detector mode (red dashed) overlaid with $\phi_{\text{B}}$ (blue solid) as a function of the position for low, $a=1/2L$ (left) and large $a=2/L$ (right) accelerations. $N=6$. The interior of the positive and negative absolute value envelopes have been shaded, and the real part is drawn inside indicating the oscillation of the wave. The black vertical line shows the position of Rob's horizon, illustrating that as acceleration increases more of $\phi_{\text{B}}$ is inaccessible to Rob. }
\end{figure}

\begin{figure}[h]
\begin{center}
\includegraphics[width=\columnwidth]{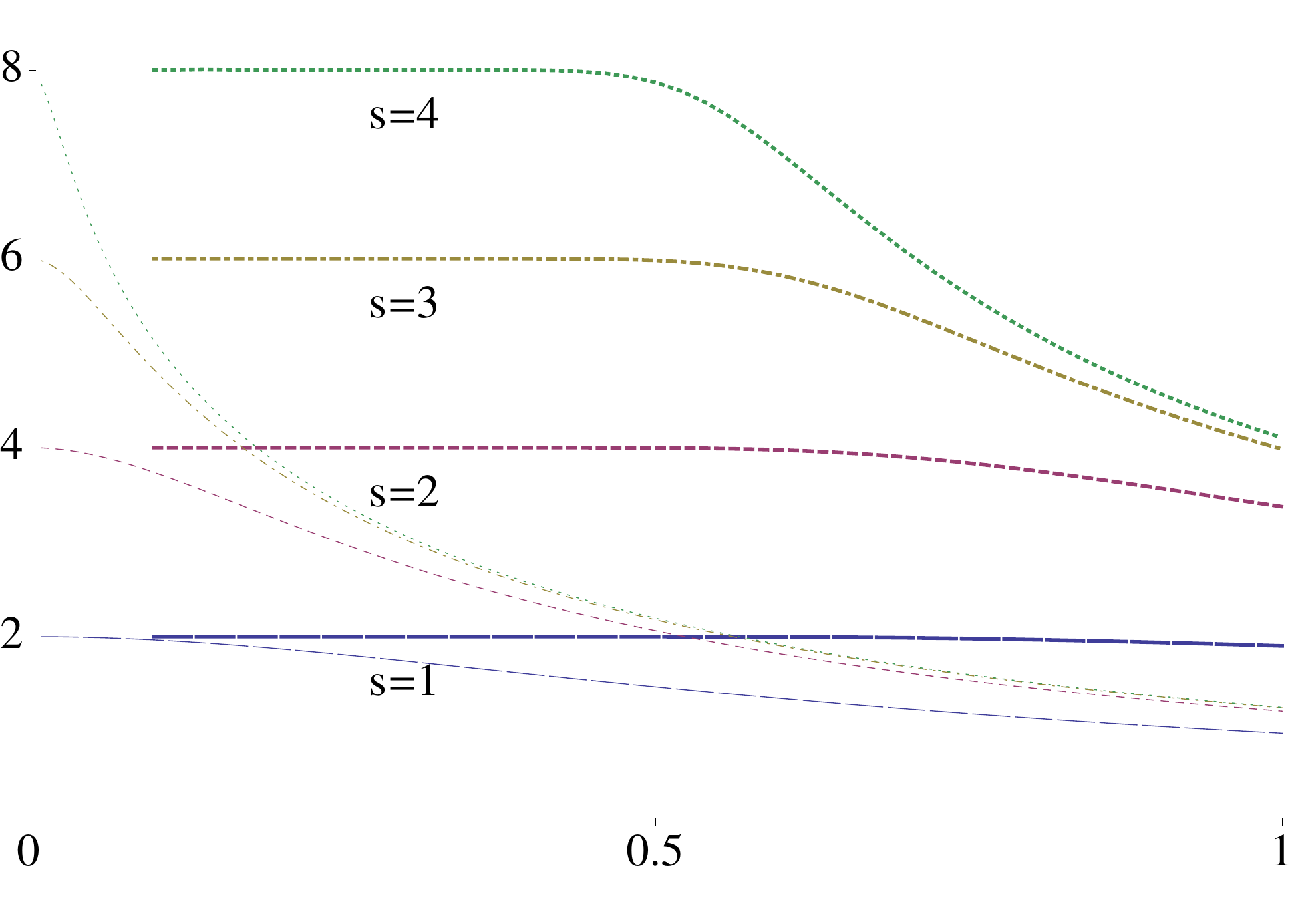}
\end{center}
\caption{\label{fig:optentanglement} Maximal amount of entanglement, $E_{\cal{N}}$  (vertical axis), available to Alice and Rob for different values of the initial squeezing, $s=1,2,3,4$ (thick lines), as a function of the dimensionless acceleration parameter $\frac{a L}{c^2}$. Also shown is a comparision with the entanglement that would have been extracted by the accelerated Gaussian detectors (thin lines) described in the previous section of the text. In all plots, $N=6$ and the detector low Rindler frequency cut-off is assumed to be $\tfrac{c}{2L}$. }
\end{figure}

\emph{--Optimized accelerating detectors.\thinspace} Consider the decomposition of Bob's mode in terms of the positive frequency Minkowski plane waves, $u_k$, and the region I and II positive frequency Rindler plane waves, $w_{Ik}$ and $w_{IIk}$, respectively:
\bea
\phi_{\text{B}}&=& \sum_k (u_{k}, \phi_{\text{B}})u_{ k} \label{minkowskidecomp}\\
&=&\sum_k(w_{Ik},\phi_{\text{B}})w_{Ik}+(w_{IIk},\phi_{\text{B}})w_{IIk}\nonumber\\&~&-(w_{Ik}^*,\phi_{\text{B}})w_{Ik}^*-(w_{IIk}^*,\phi_{\text{B}})w_{IIk}^*.\label{rindlerdecomp}
\eea
From the second line we see that this mode contains contributions from region II Rindler modes, $w_{IIk}, w_{IIk}^*$, and negative frequency region I Rindler modes, $w_{Ik}^*$. Due to the event horizon, the Region II frequencies are completely inaccessible to Rob, and therefore he can not build a detector which sees them. Likewise, the negative Rindler frequency modes in region I are also unusable in the construction of the annihilation operator associated with his detector.  
Therefore, we first try to define an optimized detector mode by simply removing these components from $\phi_B$ and renormalizing the mode:
\beq\label{phioptdef}
\psi_{opt}\equiv |N| \sum_k (w_{Ik}, \phi_{\text{B}})w_{Ik}; \quad |N|=\frac{1}{\sqrt{\sum_k|(\phi_{\text{B}},w_{Ik})|^2}}.
\eeq
A small calculation shows that $(\psi_{opt},\phi_{\text{B}})=|N|^{-1}$. Indeed, this mode is optimized in the sense that for any other accelerated detector mode in region I, $\psi'=\sum_k (w_{Ik},\psi')w_{Ik}$, the magnitude of $|\beta|=|(\psi',\phi_{\text{B}})|$ is bounded from above by $(\psi_{opt},\phi_{\text{B}})$:
\bea 
|(\psi',\phi_{\text{B}})|^2&=&|\sum_k (w_{Ik},\psi')(w_{Ik},\phi_{\text{B}})|^2\nonumber\\
&\leq& \sum_k |(w_{Ik},\psi')|^2\sum_j |(w_{Ij},\phi_{\text{B}})|^2\nonumber\\
&=&\sum_k |(w_{Ij},\phi_{\text{B}})|^2=(\psi_{opt},\phi_{\text{B}})^2.
\eea
While $\psi_{opt}$ tries as much as possible to fit to $\phi_{\text{B}}$ it never completely succeeds because of penetration of part of $\phi_{\text{B}}$ beyond Rob's horizon \footnote{It is also possible to relate the negative region I frequency contribution directly to this horizon penetration. The combination $w_{Ik}^*+e^{\frac{\pi |k| c^2}{a}}w_{IIk}$ is a superposition of pure negative frequency Minkowski waves, therefore $(w_{Ik}^*,\phi_{\text{B}})= -e^{\frac{\pi |k| c^2}{a}}(w_{IIk},\phi_{\text{B}})$.}. At low acceleration, the tail of the inertial mode penetrating the horizon is very small and so the detector becomes approximately Gaussian approximating the mode it is observing, see left figure in Fig. \ref{fig:optmodepic}. However, at large acceleration the tail of Bob's mode penetrates the horizon by a larger extent, see right of same figure, and so Rob is never able to completely reconstruct all of the state thereby leading to an inevitable loss of entanglement.
In Fig.~\ref{fig:optentanglement} we plot the maximal amount of entanglement available to Rob and Alice, calculated using the optimized detector. We compare it with the simple Gaussian model that we used in the previous section. For the range of accelerations where the spread in acceleration over the effective size of the detector is small, Rob can reconstruct nearly all of the entanglement.  It is only once the acceleration becomes large compared with $\frac{c^2}{L}$ that degradation effects begin to appear, precisely when a modest amount of the inertial mode is out of Rob's view.
We have therefore found that at large accelerations entanglement degradation can never be completely avoided.  

\emph{--Conclusion.\thinspace} We have revisited the problem of the degradation of entanglement in non-inertial frames, giving a localized and physically meaningful discussion of the problem. This has allowed us to answer the question originally posed several years ago on how entanglement is effected by accelerated motion.  We have traced the source of entanglement degradation to a mode-mismatch that will be ubiquitous in any definition of a detector mode, although some detector's can be made better than others. 

We have found that the Unruh noise did not feature in the observed degradation. It would seem in practice, that for realistic detectors of the type we consider, the Unruh noise is a sub-leading effect to the mode-loss effect we have described. This suggests that entanglement degradation would be actually much easier to observe than the Unruh effect itself which could be of a great importance for the future experimental studies of non-inertial effects on quantum phenomena.

\acknowledgments
J. D. thanks B. Munro and K. Nemoto for useful discussions.

\end{document}